\documentclass[aip, jap,preprint]{revtex4-2}
\usepackage{graphicx}
\usepackage{amsmath}
\usepackage{hyperref}

\usepackage{siunitx}
\newcommand{\NbOx}{NbO$_\mathrm{x}$}
\newcommand{\NbOt}{NbO$_\mathrm{2}$}
\newcommand{\NbtOf}{Nb$_\mathrm{2}$O$_\mathrm{5}$}
\newcommand{\degC}{$^\circ$C~}

\begin{document}

%\preprint{}

\title{Measurement of the Crystallization and Phase Transition of Niobium Dioxide Thin-Films for Neuromorphic Computing Applications Using a Tube Furnace Optical Transmission System} %Title of paper



\author{Zachary R.\ Robinson}
\affiliation{Department of Physics, SUNY Brockport, Brockport, NY}
\email[]{zrobinson@brockport.edu}

\author{Karsten Beckmann}
\affiliation{College of Nanoscale Science and Engineering, SUNY Polytechnic Institute, Albany NY}
\affiliation{NY CREATES, Albany NY}

\author{James Michels}
\author{Vincent Daviero}
\author{Elizabeth A. Street}
\affiliation{Department of Physics, SUNY Brockport, Brockport, NY}

\author{Fiona Lorenzen}
\author{Matthew C.\ Sullivan}
\affiliation{Department of Physics and Astronomy, Ithaca College, Ithaca NY}

\author{Nathaniel Cady}
\affiliation{College of Nanotechnology, Science \& Engineering, University at Albany, Albany, NY}

\author{Alexander Kozen}
\affiliation{Department of Physics, University of Vermont, Burlington, VT}

\author{Jeffrey M. Woodward}
\author{Marc Currie}
\affiliation{U.S. Naval Research Laboratory, Washington, DC}

% Collaboration name,if desired (requires use of superscriptaddress option in \documentclass). 

\date{\today}

\begin{abstract}

Significant research has focused on low-power stochastic devices built from memristive materials. These devices foster ‘neuromorphic’ approaches to computational efficiency enhancement in merged biomimetic and CMOS architectures due to their ability to phase transition from a dielectric to a metal at an increased temperature. Niobium dioxide has a volatile memristive phase change that occurs $\sim$800\degC that makes it an ideal candidate for future neuromorphic electronics. A straightforward optical system has been developed on a horizontal tube furnace for \emph{in situ} spectral  measurements as an as-grown \NbtOf\ film is annealed and ultimately crystallizes as \NbOt. The system measures the changing spectral transmissivity of \NbtOf\ as it undergoes both reduction and crystallization processes. We were also able to measure the transition from metallic-to-non-metallic \NbOt\ during the cooldown phase, which is shown to occur about 100\degC lower on a sapphire substrate than fused silica. After annealing, the material properties of the \NbtOf\ and \NbOt\ were assessed via X-ray photoelectron spectroscopy, X-ray diffraction, and 4-point resistivity, confirming that we have made crystalline \NbOt. 

\end{abstract}


\maketitle 

\section{Introduction}
Structural analyses of materials with pressure and temperature dependent crystallographic phase transitions have been studied for over fifty years \cite{peierls1955quantum,polizzi2016optical,sakata1969note,goodenough1971two,pynn1976unusual,posternak1979electronic,gervais1979infrared}. Considerable discussion regarding their underlying physical mechanism focused on Peierls versus Mott-Hubbard approaches \cite{wentzcovitchvo1994,wentzcovitch1994-2,britodynamic2017}. Specifically for niobium dioxide, the material studied in this paper, considerable evidence has shown the phase transition to be a reversible second-order Peierls transition \cite{streltsov2014orbital, wahila2019evidence}. This second order phase transition corresponds to the thermally-induced change from the body-centered tetragonal (BCT) phase to a rutile phase. \NbOt\ undergoes such a second order phase change resulting in the transition from dielectric to metallic properties with an accepted metal-insulator transition temperature at  $\sim$810\degC \cite{wahila2019evidence}. The transition itself is accompanied by an orders-of-magnitude decrease in resistivity, making the material ideal for so-called `neuromorphic' computing applications \cite{pickett2013scalable}. \\
 
A wide range of phase transition materials have been evaluated in terms of their electrical properties, as cited in many of the references above. These include several analyses focusing on \NbtOf\ and \NbOt. However, measurement of the phase transition temperature is often complicated by the fact that the electrical devices are `electroformed,' which is a non-reversible process of creating crystalline filaments out of amorphous material by applying a large voltage between electrodes \cite{sullivan2022threshold,li2018anatomy,nandi2020electric,kozen2020situ,chudnovskii1996electroforming}. The filament formation is stochastic in nature, with a significant amount of device-to-device variation in operational parameters, such as threshold and hold voltages. Once the filaments are formed, phase transitions can be initiated by `Joule heating,' where the \NbOt\ filamentary-structures are heated by passing current through them. Once the temperature of the filament reaches the phase transition temperature, the resistance of the device abruptly switches from a high to low state. With these types of structures, which are usually on the order of $\sim$100~nm in diameter, it is often impossible to accurately measure the phase transition temperature. 

Other groups have formed crystalline \NbOt\ by thermally annealing as-deposited films \cite{kozen2022crystallization,fridriksson2022growth}. This poses a number of challenges. First, the annealed films generally have quite a low resistance, such that current densities required to Joule-heat the films hot enough to initiate the phase change from insulator to conductor (BCT to rutile) are unachievable. Nanoscale patterning of the devices alleviates this problem, but is difficult and requires specialized lithographic equipment. Additionally, the temperatures required for crystallization often exceed the temperatures achievable with metal contacts on the films, given that annealing temperatures $>$800\degC are typically used \cite{kozen2022crystallization}.

Additionally, many groups have begun tuning the electrical properties of \NbOt\ devices by incorporating dopants or directly alloying the films with other transition metals, such as Ti \cite{nath2021engineering}. Measuring the impact that dopants have on the crystallization and subsequent phase change is an enormous effort, and one in which specialized synchrotron-based measurements are often required in order to extract the phase transition temperature. The phase transition temperature is a critical operational parameter for subsequent electrical devices because it is proportional to the electrical power that will be needed to heat the devices through their phase transition.

While consideration of the optical spectral transmission properties of \NbtOf\ and \NbOt\ has been made, spectral analyses on the thin film transition from dielectric \NbtOf\ to metallic \NbOt\ as the film is heated and cooled through its transition temperature is lacking from the literature. We propose it here as a straightforward technique for measuring the transition temperature as the films are engineered with strain, dopants, or in other ways that may modify the transition temperature. We believe that the system we have built will allow for fast and inexpensive analysis of uniform thin-films of \NbOt\, and may streamline the future development of doped or alloyed \NbOx films without the need for fabrication all possible stoichiometries into electrical devices. The system will also enable fundamental studies of the phase transition temperature in a straightforward manner, without the need for specialized synchrotron-based measurements. The primary purpose of this paper is to introduce the system and all of its functionality, along with results from our initial experiments. 

\section{Experimental}

The characterized films were deposited on a Kurt Lesker Inc. PVD75 argon sputtering deposition tool. Two sets of samples were fabricated on two different substrates. Optically transparent substrates were c-plane sapphire (Al$_2$O$_3$) and fused silica (FS) were used to minimize subsequent measurement interference. The deposition system is equipped with Ar and O$_2$ gases where Ar is used as the primary sputtering gas and O$_2$ added as the reactive gas enabling the deposition of oxide films. Varying the concentration of oxygen in the argon plasma allows us to achieve a wide variety of stoichiometries. This system has been used to fabricate both pure \NbOx\ films, using a Nb target, and Ti-alloyed \NbOx\ films with a  Nb0.9Ti0.1 target. This study focuses on our initial characterization of pure \NbOx\ films.

The room temperature deposition was performed at a pressure of 3~mTorr with flows of 43 and 1.3~sccm for Ar and O$_2$, respectively, resulting in a 2.9~\%~O$_2$ flow compared to the overall flow. The target RF power density for both targets was kept at 6.07~W/cm$^2$. The substrate was rotated at 10~rpm during the deposition, and the uniformity was confirmed with \emph{ex situ} ellipsometry.

X-ray Photoelectron Spectroscopy (XPS) was performed on a PHI Quantera using a monochromatic Al-K$\alpha$ X-ray source. The system can collect photoelectrons with binding energy ranging from 20 to 1000~eV. XPS data were analyzed with the CasaXPS software package. The peaks were fit with GL(30) lineshape, and doublet separation of 2.72~eV. The energy of the 5+ and 4+ oxidation states are indicated in Figure \ref{fig:xps}. Default relative sensitivity factor (RSF) values of 8.21, 2.93, and 1.0 were used for Nb3d, O1s, and C1s, respectively. All of the peak locations were normalized to the peak C1s energy, which was assigned as 285.0~eV. 

Grazing incidence X-ray diffraction (GI-XRD) was performed using a Rigaku Smartlab with 9 kW copper rotating anode source and \ang{5.0} incident and receiving soller slits. The incident angle was fixed at \ang{0.5} and 2$\theta$ was scanned from \ang{20} to \ang{75} degrees in steps of \ang{0.02}.

Resistivity measurements were made using a colinear four-point probe Cascade CPS-06 and Keithly 2400 Source Measure Unit.  In order to pierce the native oxide, we used sharp tungsten carbide tips spaced 1.59 mm apart.  We used the standard formula for thin films ($\rho = 4.52 (V/I)\cdot t $), with a correction factor of $\approx 0.8$ required for the sample size (roughly 1 cm by 1 cm).

\section{Optical Transmission Instrument}

\begin{figure}[ht]
    \centering
    \includegraphics[width=0.75\linewidth]{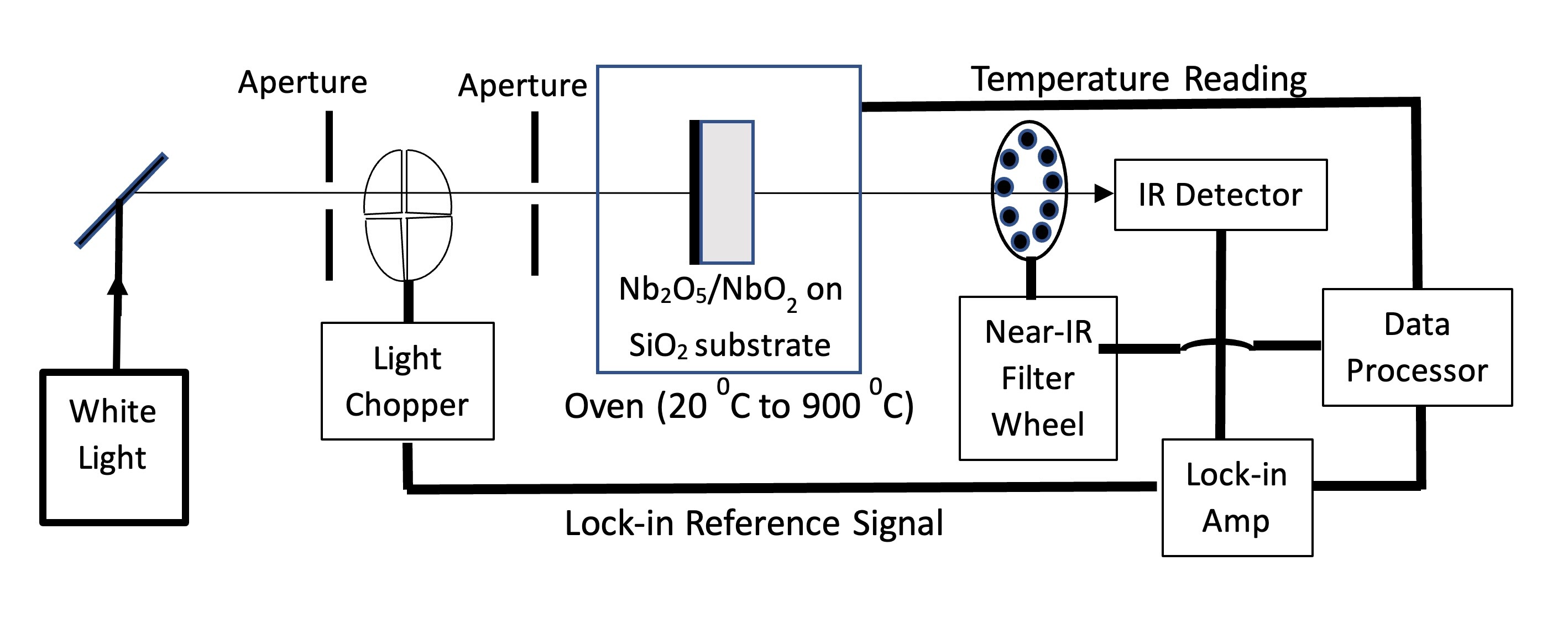}   
    \caption{Schematic of NIR transmission instrument. The light is coupled into the system from the left, passes through apertures and the chopper, is incident on the sample, and then passes into our `filter wheel spectrometer' on the right. The entire system is controlled by a Raspberry Pi data processor, along with our various other electronic components. The wavelengths measured for this study are 900nm - 1650nm.}
    \label{fig:schematic}
\end{figure}

A schematic of the near-infrared (NIR) optical system is shown in Figure \ref{fig:schematic}. It consists of a high-power stabilized quartz tungsten-halogen light source (150W, 360 - 2500 nm) with output that is passed through two apertures used for beam collimation and through the 5~cm diameter tube of an MTI Corp. oven model GSL-1500X that contains the sample. Two thin film \NbOx\ films deposited on both fused silica and sapphire substrates were used for this study. The substrates are transparent in the NIR region at the wavelengths and oven temperatures considered here (see supplemental data). The incident beam passes into the furnace through a quartz window, through the sample, which has an aperture of 0.3~in in front of and behind the sample, and out of the furnace through a second quartz window (see supplemental data for mechanical drawing of the sample holder). Finally, the light passes through one of the NIR optical bandpass filters mounted in a software controlled rotatable wheel. The beam intensity is measured with a Thorlabs model PDA50B2 IR detector with an 800~–~1800~nm range, which is placed at the focal length of a 1~in OD lens that is 1~in from the detector. Nine filters at wavelengths in the range of 900 nm to 1650 nm were used with maximum filter bandwidths of 12.0~$\pm$~2.4~nm, FWHM.

The incident beam was modulated by a light chopper operated at around 215Hz that provides a reference signal to a Stanford Research Systems model SR810 DSP lock-in amplifier. Data consisting of the lock-in amplifier output optical signal, the filter wavelength value, and the oven temperature were collected in the data processor. The oven temperature was varied at a rate of 10 \degC/min during heating and cooling. Data collection at each temperature level was collected within 45~s ($sim$5s per filter), which allowed sufficient time for the signal to stabilize between wavelengths (described in more detail below).

The data processor consists of a Raspberry Pi model 3b control unit, which records all of the data and controls the motor position using a custom Python script, which is available along with the circuit schematics and CAD files for this system. The Raspberry Pi is controlled via secure shell by a secondary computer (actually, another Raspberry Pi). The user enters the temperature into the terminal, and the controller records the output of the lock-in amplifier. The Raspberry Pi then rotates the optical filter wheel to the next filter by sending a pulse sequence to a stepper motor controller, delays 3~-~5~s to allow the lock-in amplifier to settle, and then records the next filter wheel's optical transmission data. In this way, we can perform ``filter wheel spectroscopy'' and quickly measure small signals with a wavelength range of 900nm-1650nm. One benefit of such a system is the relatively large signal that we are able to achieve, since our filters are 1~inch in diameter and the full optical signal is then focused onto the detector with a 1~in focal length lens.

Prior to operation of the system, we measured each fixed wavelength for 120~s on 1~s intervals in order to estimate the noise in our optical system. The tube furnace was kept at room temperature during this measurement. The plot of intensity vs. time can be found in the Supplemental section. Across all of the wavelengths, the standard deviation of the optical signal was less than 0.5\% of the mean value. 

Calibration of this system was performed by inserting into the tube furnace each of the blank substrates that was later used for \NbOx\ deposition. In this study, we compared the optical signal and annealing conditions for \NbOx\ films deposited on 1cm~x~1cm fused silica and sapphire substrates that were polished on both front and back surfaces. For the calibration anneals, the entire system was operated under conditions that we previously found to result in fully crystallized \NbOt\ films \cite{kozen2020situ,kozen2022crystallization}. The furnace was heated to 900\degC under a flow of 60~sccm of forming gas (5\%H$_2$, 95\% N$_2$). Over the full wavelength range, optical transmission data for the substrates (without \NbOx\ films) were collected every 50\degC, and can be found in supplemental data.

Since both substrates are approximately optically flat over the temperature range that we required for subsequent \NbOt\ crystallization, measurements on samples with thin \NbOx\ films were divided by the appropriate average value from the calibration, in order to remove the effect of the system. In other words, $\textnormal{intensity} = \frac{\textnormal{measurement signal}}{\textnormal{average calibration signal}}$.  Our analysis does not account for the small temperature dependent changes in the substrate's transmission. In our estimation, we were able to mitigate any influence from source brightness, window absorption or reflection, and detector sensitivity. 

\section{Results and Discussion}

The relative transmittance data for the thin film \NbOx\ films deposited on fused silica and sapphire substrates are presented here.
Following both anneals, the samples were removed from the tube furnace and shipped to SUNY Albany, along with un-annealed control samples, for characterization with XPS and XRD.

\begin{figure}[ht]
    \centering
    \includegraphics[width=0.85\linewidth]{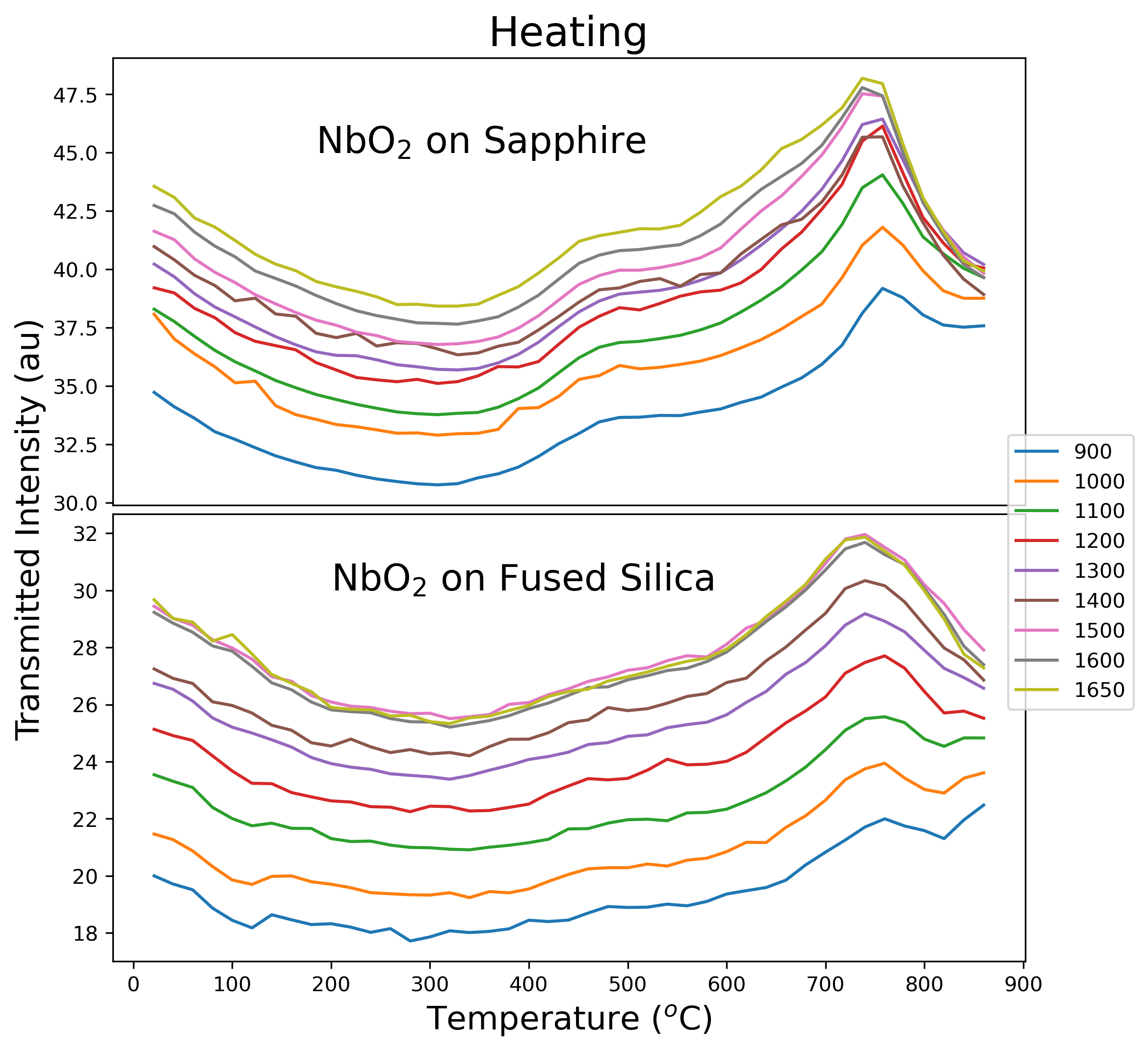}   
    \caption{Spectral optical transmission during the heating phase of the anneal, as measured on PVD-grown \NbOx\ deposited on both c-plane sapphire (upper) and fused silica (lower) substrates.}
    \label{fig:heating}
\end{figure}

Figure \ref{fig:heating} shows the relative transmittance of the \NbtOf-to-\NbOt\ anneal on both substrates versus increasing anneal temperature. During the initial heating, both samples are characterized by a near quadratic transmittance change in the temperature range from room temperature to $\sim$500\degC, reaching a minimum at $\sim$350\degC and with monotonically decreasing transmittance as a function of decreasing wavelength.

The decrease in transmittance between room temperature to $\sim$350\degC appears to signify the result of \NbtOf\ structural heating effects. Thereafter, both transmittance levels rise, somewhat linearly for the fused silica substrate, but with a plateau between 460 – 560\degC for the sapphire substrate sample. The transmittance increase in this range is likely due to removal of impurities, or the beginning stages of grain growth and coalescence in the \NbtOf. This is followed by further gradual increases peaking at 760\degC for sapphire substrate, and 740\degC for the fused silica substrate samples. We speculate that the peak corresponds to crystallization of the \NbtOf\ phase of the samples. This is based on prior work, in which we have found that similar samples (grown by atomic layer deposition) annealed to temperatures around 700\degC are crystalline \NbtOf \cite{kozen2022crystallization}.

At temperatures between 740–860\degC, several important observations can be made. First, for the longer wavelengths in the range (1500–1650)~nm, the transmittance decreases much more significantly ($\sim$20\%) for both materials compared to the shorter wavelengths. It is also interesting to note that above $\sim$830\degC, the 1400~nm wavelength shows a more significant transmittance drop than the other wavelengths, possibly due to the formation of water vapor that may evolve from the film, which has an absorption peak around 1380~nm.

Next, both samples show a spread of transmittance with respect to wavelength over the temperature range, reaching a minimum at 900\degC for the fused silica, and 860\degC for the the sapphire. We speculate that the decrease described here corresponds to initial stages of the reduction of \NbtOf\ into \NbOt, which apparently occurs more rapidly, and at a lower temperature, on the well-ordered sapphire substrate. 
   
\begin{figure}[ht]
    \centering
    \includegraphics[width=0.85\linewidth]{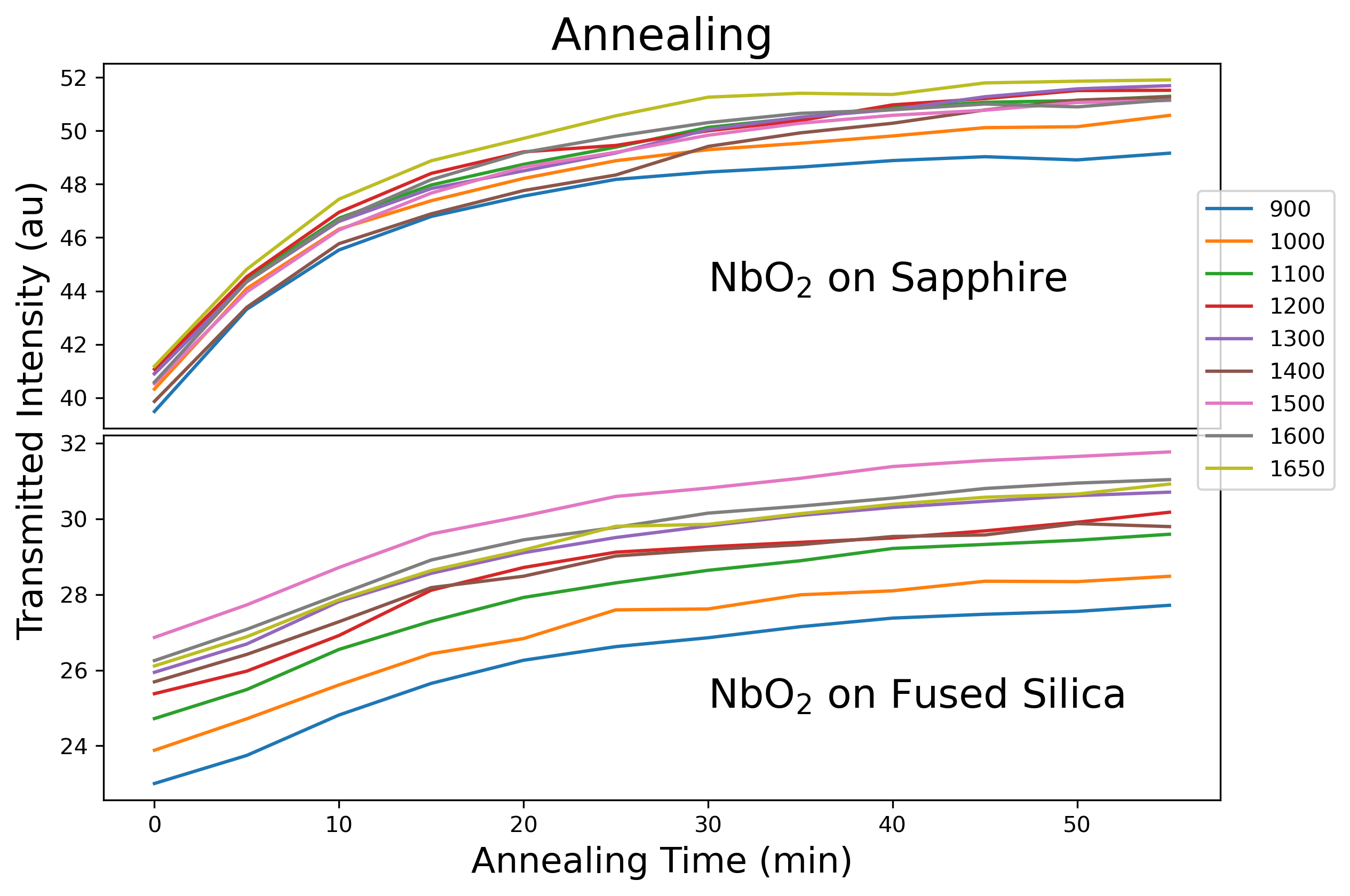}   
    \caption{Spectral optical transmission during the constant-temperature annealing phase of the anneal, as measured on PVD-grown \NbOx\ deposited on both c-plane sapphire (upper) and fused silica (lower) substrates.}
    \label{fig:anneal}
\end{figure}

The next phase of the annealing process consisted of a 60 minute hold at 900\degC. The relative transmittance plots are shown in Figure \ref{fig:anneal}. Both annealing plots indicate a process that saturates around 40 minutes, which seems to imply the reduction and crystallization processes, from as-deposited amorphous \NbtOf\ to crystalline \NbOt\ has completed, and subsequent annealing may not have any dramatic effects on the resulting films. This is consistent with some of our prior work showing that similar films are fully crystallized following a 40min anneal at 900\degC \cite{kozen2022crystallization}.

\begin{figure}[ht]
    \centering
    \includegraphics[width=0.85\linewidth]{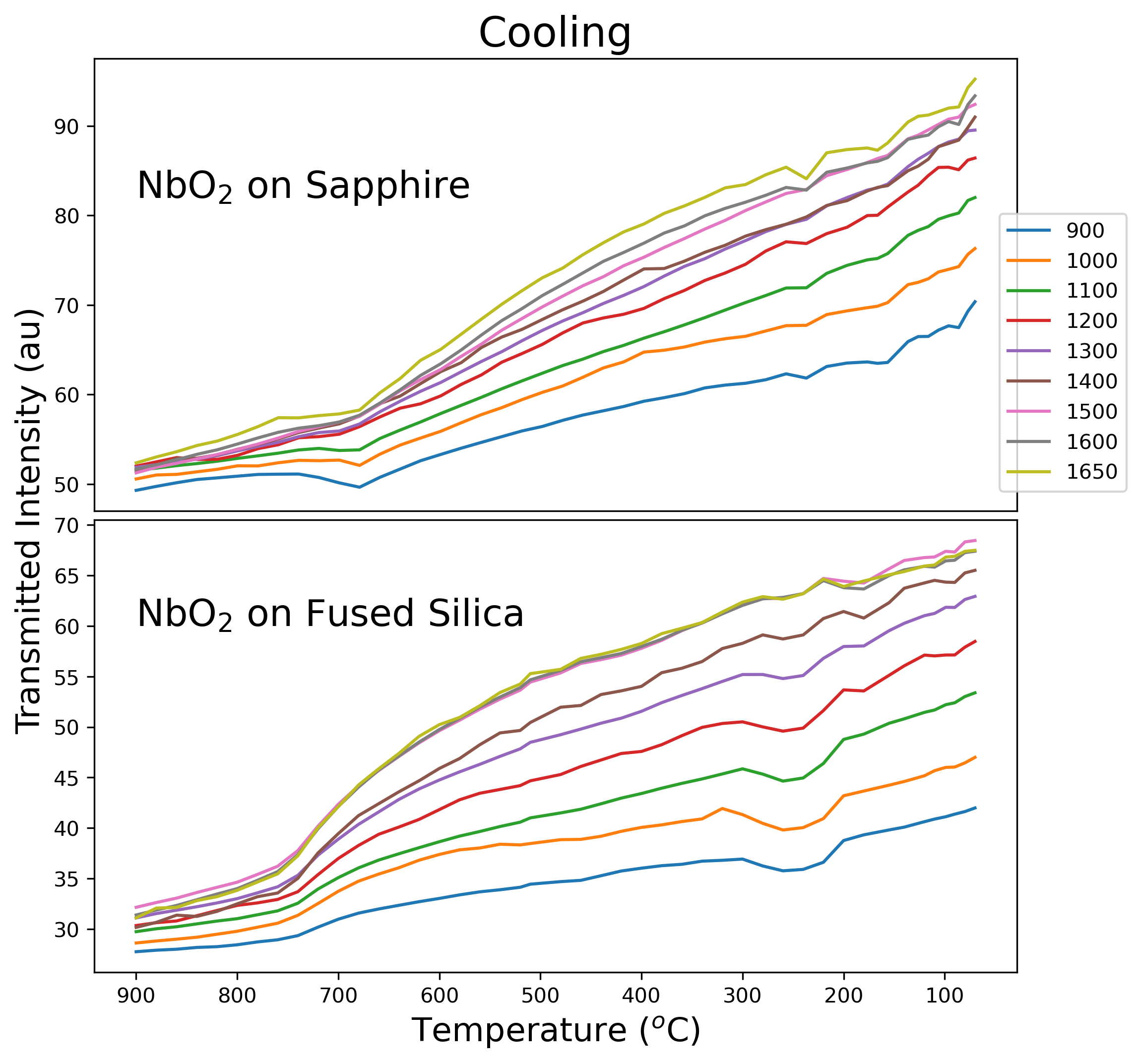}   
    \caption{Spectral optical transmission during the cooling phase of the anneal, as measured on PVD-grown \NbOx\ deposited on both c-plane sapphire (upper) and fused silica (lower) substrates.}
    \label{fig:cooling}
\end{figure}

During the post-anneal cooldown phase, which are shown in Figure \ref{fig:cooling}, increasing spectral dispersion below the inflection temperature can be observed. For the sapphire substrate sample, the inflection occurs at 680\degC and at 760\degC for the fused silica substrate sample. We speculate that the inflection may correspond to the phase transition between the insulating and conducting states, which typically occurs around 810\degC. There are several possible reasons for the discrepancy in temperature, including possible thermal hysteresis in the furnace, which may provide an offset between the furnace temperature and the film temperature. We expect that an offset between sample and furnace temperature is the same for both the fused silica and sapphire samples. It is also possible that there is a significant influence from the interaction between the \NbOt films and the underlying substrate. Substrate effects are discussed below in the context of the \emph{ex situ} XRD measurements. All of the raw data for the entire annealing process are shown in the Supplemental Data section.

After being removed from the furnace, transmittance and reflectance measurements were made at room temperature on as-grown and annealed films, see Figure \ref{fig:TR}. While the as-grown films on both substrates show similar reflectance to the annealed films, the transmittance of annealed films is much larger. This shows a dramatically reduced absorption in the annealed films. This is similar to larger NIR absorption seen in amorphous vs. crystalline semiconductors (e.g., Si and Ge) \cite{Cardona2010fundamentals}.

\begin{figure}[ht]
    \centering
    \includegraphics[width=0.85\linewidth]{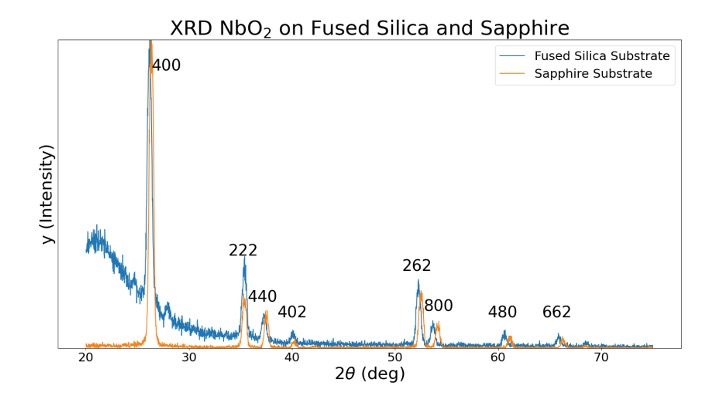}   
    \caption{XRD data for annealed \NbOx\ films, showing that the films on fused silica (upper) is approximately a powder patter with no preferred crystallographic orientation and the films on c-plane sapphire (lower) are textured.}
    \label{fig:xrd}
\end{figure}

\begin{figure}[ht]
    \centering
    \includegraphics[width=0.85\linewidth]{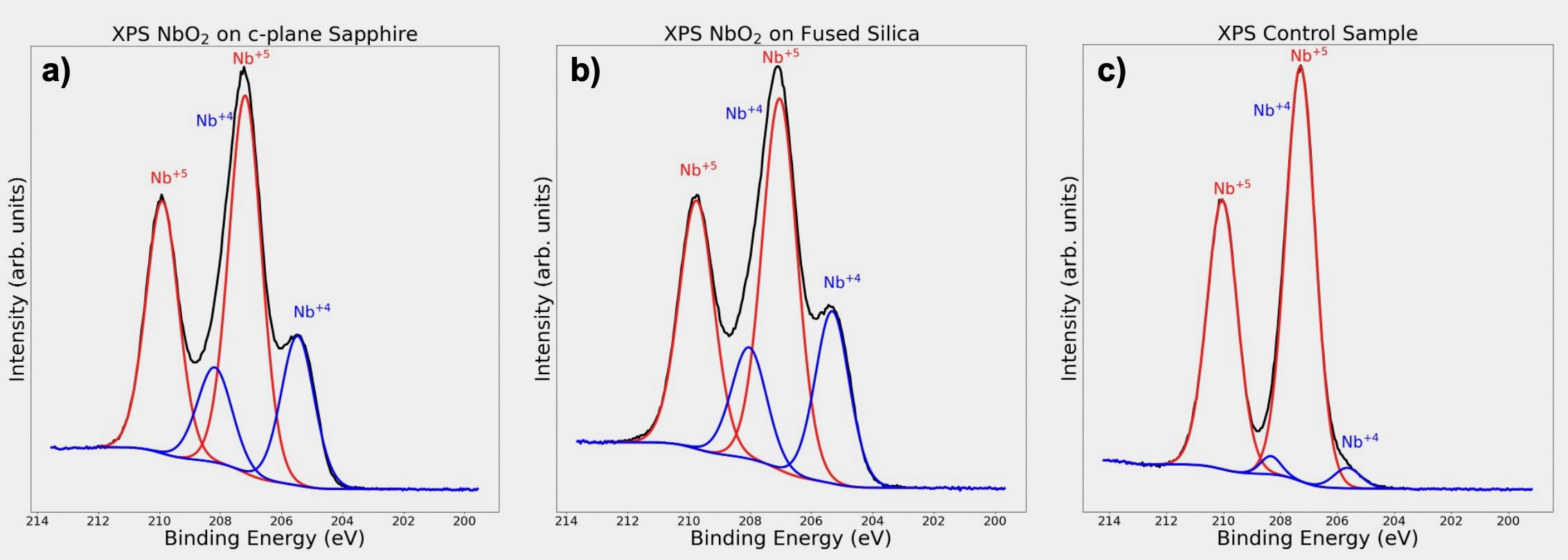}    
    \caption{XPS data for annealed \NbOx\ films on a) c-plane sapphire, b) fused silica, and c) unannealed control sample. The measurements on both annealed samples show the typical \NbOt\ spectra, while the control shows a typical spectra for \NbtOf\ grown with our 3\% oxygen flow process.}
    \label{fig:xps}
\end{figure}

\begin{figure}[ht]
    \centering
    \includegraphics[width=0.85\linewidth]{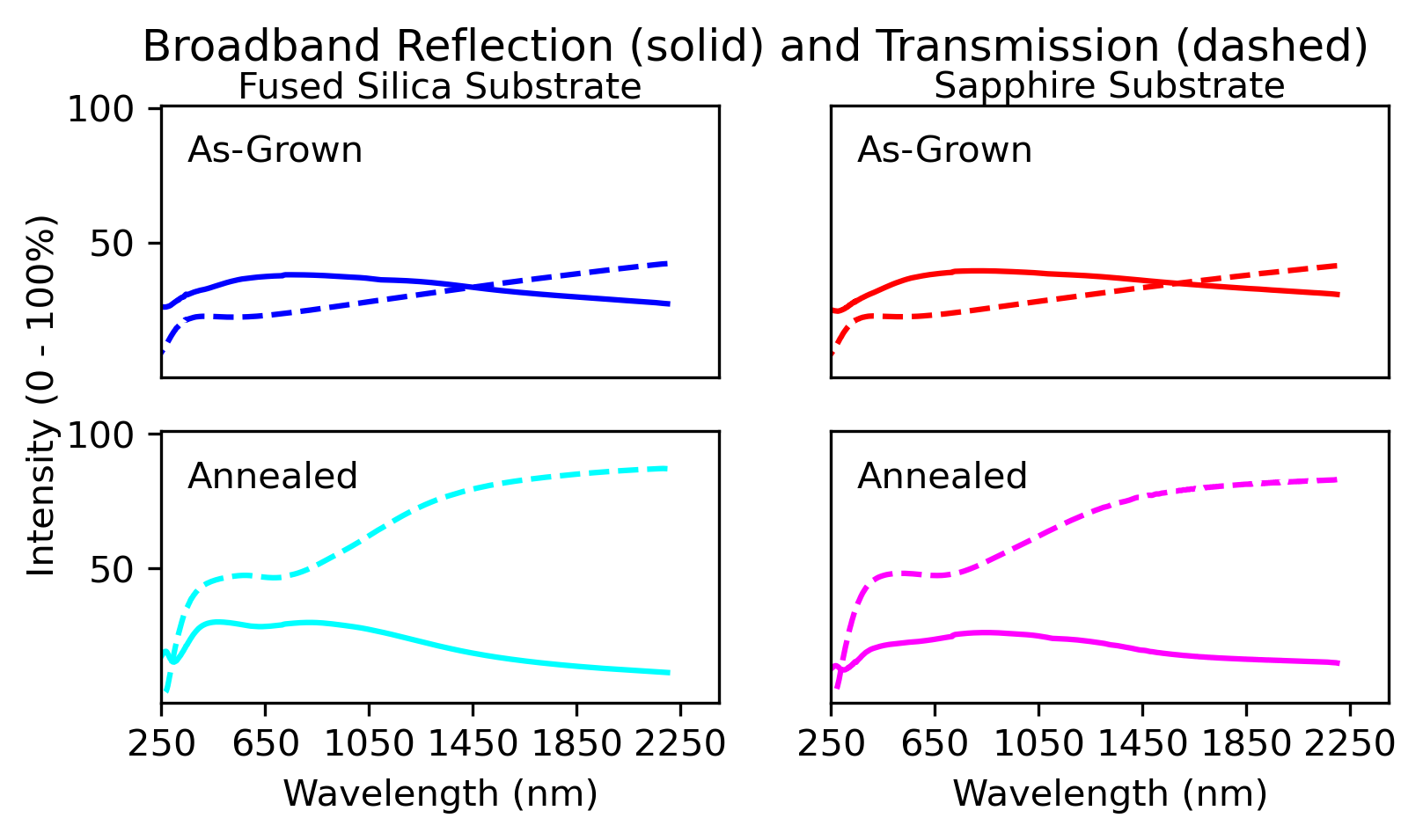}    
    \caption{Transmittance and Reflectance of as-grown and annealed NbOx thin films on fused silica and sapphire substrates. There is a noticeable increase in transmittance after annealing (on both substrates) with little change in the reflectance. This shows a much stronger absorption by the as-grown films in the near-IR region.}
    \label{fig:TR}
\end{figure}

On the fused silica substrate, XRD measurements show that the film very closely approximates a powder diffraction pattern (see Figure \ref{fig:xrd}). This indicates that the substrate-film interaction may be quite weak, such that there is no preferred orientation between the film and the underlying substrate.

On sapphire, a stronger interaction between the substrate and film has been observed.  This is apparent in the orange XRD plot, for which the relative intensity of the (400) peak compared to the other peaks is greater than that observed for the film on the fused silica substrate, indicating an increase in preferential orientation of the crystalline structure.  Note that the absence of the amorphous feature at 2$\theta~\sim$~\ang{21}, which has been previously observed in as-grown \NbtOf (see ref \cite{kozen2022crystallization}), may be the result of a lower critical angle compared to the film on fused silica (see supplemental figure 3). The lower critical angle suggests that the GIXRD measurement is less surface sensitive, and therefore diminishing the 

Given that the films deposited on sapphire appear to have a much stronger interaction with the underlying substrate, we therefore speculate that the phase transition temperature (that is, the switch between the low resistance and high resistance phase of \NbOt) may be affected. This effect has been observed in VO$_\textup{2}$, which undergoes a similar phase transition in electrical and optical properties (albeit at a much lower temperature). In VO$_\textup{2}$, measured transition properties vary widely and can depend on growth technique, crystal size, annealing, doping, strain, and other factors. In addition, the transition is hysteretic, occurring at different temperatures when heating vs. cooling. Crystal facets/texture, vacancies, grain size, and stress have been correlated with variations in transition hysteresis width \cite{Currie2019asymmetric, goodenough1971two, Case_1984}. Therefore, using the substrate to effect a change in the phase transition temperature is a powerful technique for engineering carefully designed \NbOt\ films that transition between phases at specific temperatures. Ongoing experiments aim to study the influence of the substrate on the phase transition temperature in more detail.

In addition to XRD, the films were also characterized with XPS. XPS measurements were taken on a control films deposited on an identical substrates simultaneously with the films that were annealed, and also on the annealed and fully crystallized \NbOt\ films after they were removed from the tube furnace. There is only one control film shown since the fused silica and sapphire had essentially identical spectra prior to anneal. The XPS data can be found in Figure \ref{fig:xps}. The data reflect what we typically observe in XPS data for fully crystallized \NbOt\ films - a primary doublet peak corresponding to a $\sim$1.5~nm native oxide of \NbtOf, and a secondary set of peaks corresponding to a buried \NbOt. The relatively intense \NbtOf\ peaks result from atmospheric exposure of the films, as we previously reported \cite{kozen2022crystallization}. Growth of the +4 oxidation state peaks following anneal confirms what we see in XRD, indicating that we have a fully crystallized stoichiometric \NbOt film with a thin native \NbtOf oxide in the surface region.

Resistivity measurements were also performed, as can be found in Table \ref{tab:resistivity}. For both the fused silica and sapphire substrate, the resistivity of the films increases following the reduction/crystallization anneal by about 4 orders of magnitude. The annealed \NbOt\ films have a resistivity similar to what has been reported in the literature, whereas the as-grown \NbtOf is many orders of magnitude more conducting than what has been reported \cite{stoever2020approaching}. The decreased resistivity in the as-grown materials, compared to the literature, are likely caused by impurities in the film that are removed during the anneal. 

\begin{table}
\begin{tabular}{|c|c|c|}
    \hline
    Sample & Resistivity ($\Omega$-cm) & Standard Deviation\\\hline
   Sapphire (control)  & 0.009 & 0 \\\hline
   Sapphire (annealed) & 156 & 2\\\hline
   Fused Silica (control) & 0.009 & 0 \\\hline
   Fused Silica (annealed) & 196& 4\\\hline
\end{tabular}
    \caption{Table of resistivity values for the unannealed and annealed \NbOx\ films. We estimate a 10\% uncertainty on all of the measurements, based on the estimated uncertainty in the film thickness (nominal film thickness is 40nm). The standard deviation is a result of 5 measurements from 5 different locations, and implies that our films (control and annealed) are very uniform.}
    \label{tab:resistivity}
\end{table}

\begin{acknowledgments}
This research was supported by the National Science Foundation grants nos. DMR-2103197, DMR-2103185 and the Air Force Research Laboratory grant \#1152303-1-83972. 
M.C. acknowledges funding from the Office of Naval Research.

\end{acknowledgments}

\pagebreak
\bibliography{Bibliography}
\pagebreak

\section{Figures}

\end{document}